\pdfoutput=1

\documentclass{article}
\usepackage{ijcai20}

\usepackage{times}

\usepackage{soul}
\usepackage{url}
\usepackage[hidelinks]{hyperref}
\usepackage[utf8]{inputenc}
\usepackage[small]{caption}
\usepackage{graphicx}
\usepackage{amsmath}
\usepackage{booktabs}
\usepackage{balance}
\usepackage{algorithmic}
\usepackage{algorithm}

\usepackage{tikz}
\usetikzlibrary{calc, positioning, arrows, shapes.multipart, intersections}
\usepackage{multirow}
\usepackage{threeparttable}
\usepackage{diagbox}
\usepackage{array}
\newcolumntype{R}[1]{>{\raggedleft\let\newline\\\arraybackslash\hspace{0pt}}m{#1}}
\title{Relational Thematic Clustering with Mutually Preferred Neighbors}

\author{
Tiantian He$^1$\footnote{Contact Author}\and
Lu Bai$^1$\and
Yew-Soon Ong$^{1}$\\
\affiliations
$^1$School of Computer Science and Engineering, Nanyang Techonological University\\
\emails
\{tiantian.he, bailu, asysong\}@ntu.edu.sg
}

\begin{document}

\maketitle

\begin{abstract}
Automatically learning thematic clusters in network data has long been a challenging task in machine learning community.
A number of approaches have been proposed to accomplish it, utilizing edges, vertex features, or both aforementioned.
However, few of them consider how the quantification of dichotomous inclination w.r.t. network topology and vertex features may influence vertex-cluster preferences, which deters previous methods from uncovering more interpretable latent groups in network data.
To fill this void, we propose a novel probabilistic model, dubbed Relational Thematic Clustering with Mutually Preferred Neighbors (RTCMPN).
Different from prevalent approaches which predetermine the learning significance of edge structure and vertex features, RTCMPN can further learn the latent preferences indicating which neighboring vertices are more possible to be in the same cluster, and the dichotomous inclinations describing how relative significance w.r.t. edge structure and vertex features may impact the association between pairwise vertices.
Therefore, cluster structure implanted with edge structure, vertex features, neighboring preferences, and vertex-vertex dichotomous inclinations can be learned by RTCMPN.
We additionally derive an effective Expectation-Maximization algorithm for RTCMPN to infer the optimal model parameters.
RTCMPN has been compared with several strong baselines on various network data. 
The remarkable results validate the effectiveness of RTCMPN.
\end{abstract}

\section{Introduction}
Relational data, e.g., friends on social networking sites and cited documents in scientific corpora, are ubiquitous in the real world.
Network, which contains a set of vertices and edges, respectively representing data samples and sample-sample relations, is thereby ideal for preserving illustrative information on the data samples and complex sample-sample relationships.
Due to its universality, how to effectively analyze the network data has been of great importance.
Among various analytical tasks on network data, cluster analysis is one of the most important, and it is directly related to many real-world applications, such as social group detection~\cite{he2019contextual}, social recommendation~\cite{yang2013}, biological module discovery~\cite{airoldi2008mixed}, and topic modeling in scientific articles~\cite{chang2009}.

Cluster analysis in networks has been a long-lasting and challenging problem in machine learning community.
A number of approaches, which are either heuristic or model-based, have been proposed to effectively uncover the network clusters, by feat of maximizing the group-wise vertex cohesiveness regarding edge density, vertex features, or both aforementioned. 
For example, modularity maximization~\cite{clauset2004}, which aims at optimizing the difference in terms of density of vertices in the same group and that if the vertices are randomly grouped, is a prevalent heuristic measure for cluster analysis, and it has been incorporated into many approaches to network clustering, e.g., Fast unfolding algorithm~\cite{blondel2008}.
Besides, modern machine learning techniques, including spectral analysis~\cite{shi2000,wang2019gmc}, matrix factorization (MF)~\cite{yang2013,wang2016semantic,ye2018deep,xu2019gromov}, and probabilistic modeling~\cite{chang2009,peng2015scalable,bojchevski2018bayesian}, have also been used to build effective model-based approaches to immediately uncover network clusters.
Not only utilizing edge structure, but also incorporating vertex features, model-based approaches can unfold network clusters and simultaneously improve the interpretability of them by summarizing their themes.
Thus, many recent methods attempt to adapt cutting-edge learning techniques to discover clusters in various network data.

Though very effective in uncovering network clusters via simultaneously learning in edge structure and features, most prevalent approaches perform the task only relying on observed data, and have to predefine the learning significance of structure and features.
They inevitably overlook the latent neighbor preference indicating which proximal vertices are more possible to be in the same cluster, and hidden dichotomous inclination describing how relative significance of network topology and vertex features may trigger a pair of vertices to be related.
To fill this void, in this paper, we propose a novel probabilistic model, dubbed Relational Thematic Clustering with Mutually Preferred Neighbors (RTCMPN), to learn vertex-cluster membership, concerning not only edge structure and features, but also the latent neighborhood constraints and corresponding vertex-wise dichotomous inclination about topology and features.
The contributions of this paper can be summarized as follows:
\begin{itemize}
	\item We propose RTCMPN, which is a novel probabilistic model for revealing clusters in the network data. RTCMPN is able to learn the latent neighboring preferences, that indicate which proximal vertices have the same cluster membership, considering the vertex-wise dichotomous inclination w.r.t. network topology and features. Modeling the associations between vertex-cluster membership and the mentioned latent neighboring preference, edge structure, and vertex features, RTCMPN is capable of learning more meaningful clusters in the network data.
	\item For RTCMPN, we design a novel generative process for generating network data, based on which, we formulate the clustering task as optimizing a unified likelihood function. We also derive a novel Expectation-Maximization algorithm for learning the optimal variables, and parameters of the model.
	\item RTCMPN has been extensively compared with a number of strong baselines on various network datasets. The experimental results show that RTCMPN outperforms all the baselines on most datasets, which validate the effectiveness of the proposed model.
\end{itemize}

The rest of this paper is organized as follows. In Section~\ref{related-works}, the previous works related to the proposed model are investigated. In Section~\ref{model}, we elaborate the proposed RTCMPN, derive the EM algorithm for learning the model parameters, and analyze the computational complexity of the proposed model. The extensive experiments that are used to verify the effectiveness of RTCMPN are presented in Section~\ref{exp}. In the last section, we conclude the paper and discuss future works. 

\section{Related works}\label{related-works}
To effectively discover clusters in network data, a number of approaches have been proposed.
Some of them are able to uncover clusters utilizing network topology.
For example, Clauset-Newman-Moore algorithm~\cite{clauset2004}, Stochastic block model (SBM)~\cite{airoldi2008mixed,peng2015scalable}, Deep autoencoder-like nonnegative matrix factorization (DANMF)~\cite{ye2018deep}, Modularized nonnegative matrix factorization (M-NMF)~\cite{wang2017community}, and Normalized cut (Ncut)~\cite{shi2000} are widely used approaches, which perform the task of network clustering utilizing either edge structure or topological similarities.

More recent methods attempt to discover clusters as well as learn the descriptive features of them.
For example, Relational topic models~\cite{chang2009}, Communities from edge structure and node attributes (CESNA)~\cite{yang2013}, Semantic community identification (SCI)~\cite{wang2016semantic}, Adaptive semantic community detection (ASCD)~\cite{qin2018adaptive}, Contextual correlations preserving graph clustering~\cite{he2019contextual}, Attributed Markov random filed model~\cite{he2019end}, and Multi-view spectral clustering~\cite{kumar2011co,wang2019gmc}, are prevalent approaches, which perform the task of network clustering by incorporating edge structure with vertex features.
It is observed that none of the previous methods for network clustering considers simultaneously modeling latent neighborhood constraints and corresponding vertex-wise dichotomous inclination about topology and features.
This motivates us to propose a novel model to fill the void.

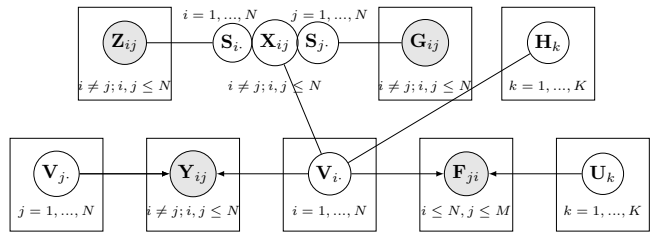
\begin{figure}[t]
	\centering
	\tikzstyle{mycir} = [circle, draw,  inner sep=2pt, text centered]
	\tikzstyle{myrec} = [rectangle, draw, minimum width=1.7cm, minimum height = 1.7cm]
	\tikzset{>=latex}
	\centering
	\resizebox{\linewidth}{!}{
	\begin{tikzpicture}[->]
	
	\node [mycir] (c1) at (0,-1.2) {$\mathbf V_{j\cdot}$};
	\node [myrec] (r1) at (0,-1.4) {};
	\node (t1) at (0, -1.9) {\scriptsize $j=1,...,N$}; 
	
	
	\node [mycir] (c2) at (4,1.2) {$\mathbf X_{ij}$};
	\node (t2) at (4, 0.4) {\scriptsize $i\ne j; i,j\leq N$};
	
	\node [mycir] (c7) at (4.78,1.2) {$\mathbf S_{j\cdot}$};
	\node (t7) at (5, 1.7) {\scriptsize $j=1,...,N$};
	
	\node [mycir] (c8) at (3.22,1.2) {$\mathbf S_{i\cdot}$};
	\node (t8) at (3, 1.7) {\scriptsize $i=1,...,N$};
	
	\node [mycir, fill=gray!20] (c9) at (1.25,1.2) {$\mathbf Z_{ij}$};
	\node [myrec] (r9) at (1.25,1) {};
	\node (t9) at (1.25, 0.4) {\scriptsize $i\ne j; i,j\leq N$};
	
	\node [mycir, fill=gray!20] (c10) at (6.75,1.2) {$\mathbf G_{ij}$};
	\node [myrec] (r10) at (6.75,1) {};
	\node (t10) at (6.75, 0.4) {\scriptsize $i\ne j; i,j\leq N$};
	
	\node [mycir] (c11) at (9,1.2) {$\mathbf H_{k}$};
	\node [myrec] (r11) at (9,1) {};
	\node (t11) at (9, 0.4) {\scriptsize $k=1,...,K$};
	
	\node [mycir, fill=gray!20] (c3) at (2.5,-1.2) {$\mathbf Y_{ij}$};
	\node [myrec] (r3) at (2.5,-1.4) {};
	\node (t3) at (2.5, -1.9) {\scriptsize $i\ne j; i,j\leq N$};
	
	\node [mycir] (c4) at (5,-1.2) {$\mathbf V_{i\cdot}$};
	\node [myrec] (r4) at (5,-1.4) {};
	\node (t4) at (5, -1.9) {\scriptsize $i=1,...,N$};
	
	\node [mycir, fill=gray!20] (c5) at (7.5,-1.2) {$\mathbf F_{ji}$};
	\node [myrec] (r5) at (7.5,-1.4) {};
	\node (t5) at (7.5, -1.9) {\scriptsize $i\leq N, j\leq M$};
	
	\node [mycir] (c6) at (10,-1.2) {$\mathbf U_{ k}$};
	\node [myrec] (r6) at (10,-1.4) {};
	\node (t6) at (10, -1.9) {\scriptsize $k=1,...,K$};
	
	\draw[] (c1)--(c3);
	\draw[-] (c2)--(c4);
	\draw[] (c4)--(c3);
	\draw[] (c1)--(c3);
	\draw[] (c4)--(c5);
	\draw[] (c6)--(c5);
	\draw[-] (c11)--(c4);
	\draw[-] (c8)--(c9);
	\draw[-] (c7)--(c10);
	\end{tikzpicture}}
	\caption{Graphical representation of Relational Thematic Clustering with Mutually Preferred Neighbors. Shaded and blank circles represent data and latent variables, respectively.}\label{graphical-model}
\end{figure}

\section{Relational thematic clustering with inclination aware neighbors}\label{model}
In this section, we elaborate the proposed RTCMPN. First, we introduce the mathematical notations. The model structure is then introduced according to the newly designed generative process. At last, we derive an EM algorithm to infer the optimal parameters of RTCMPN and analyze the computational complexity of the model.

\subsection{Notations}
Given a network composed of $N$ vertices, $\mid$E$\mid$ edges, $M$ vertex features, and $K$ ground-truth clusters, we use two binary matrices $\mathbf Y \in \{0,1\}^{N \times N}$ and $\mathbf {F} \in \{0,1\}^{M \times N}$ to represent whether two vertices are connected and whether a vertex has a corresponding feature, respectively.
We denote $\mathbf Z \in R_{+}^{N \times N}$ and $\mathbf G \in R_{+}^{N \times N}$ as the observed vertex-vertex similarities in terms of vertex topology and its features, respectively.
In this paper, cosine similarity is adopted to compute the mentioned likeness.
The $(i,j)$-th element, $i$-th row, and $j$-th column of a matrix $\mathbf Y$ are denoted as $\mathbf Y_{ij}$, $\mathbf Y_{i\cdot}$, and $\mathbf Y_{j}$, respectively.
To build the proposed model, we use $K$, $M$, and $N$ multinomial parameters $\mathbf V_{i\cdot}$, $\mathbf U_{k}$, and $\mathbf H_{k}$ to respectively represent the cluster preference of vertex $i$, the feature-theme proportion for cluster $k$, and the vertex-proportion for cluster $k$.
We use $N$ dimensional multinomial variables $\mathbf X_{i\cdot}$ and binomial variables $\mathbf S_{i\cdot}$ to represent latent neighbor preference and dichotomous inclination of topology and features of vertex $i$.

\subsection{The generative process}\label{structure}
RTCMPN takes into account the modeling of edge structure, vertex features, latent neighbor preference, and dichotomous inclination. Therefore, we design a novel generative process for the model, and the corresponding graphical representation is depicted in Fig.~\ref{graphical-model}. Given a set of network data, RTCMPN generates them according to the process shown as follows.
\begin{itemize}
	\item For each vertex $i$, draw $K$ and $N$ dimensional multinomial parameters $\mathbf V_{i\cdot}$, $\mathbf X_{i\cdot}$, and binomial parameters $\mathbf S_{i\cdot}$ as cluster preference, latent neighbor preference, and dichotomous inclinations;
	\item For each cluster $k$, draw $M$ and $N$ dimensional multinomial parameters $\mathbf U_{k}$ and $\mathbf H_{k}$ as cluster theme and cluster-vertex proportion;
	\item For each pair of vertices, draw $\mathbf X_{ij}$,$\mathbf S_{i}$,$\mathbf S_{j}$,$\mathbf Z_{ij}$,$\mathbf G_{ij}\!\sim\!Boltzmann(-\epsilon_{ij} |C)$, where $\epsilon_{ij} = -\mathbf X_{ij}(\mathbf S_{i1}\mathbf S_{j1} \mathbf Z_{ij}+\mathbf S_{i2}\mathbf S_{j2}\mathbf G_{ij})$; draw $\mathbf Y_{ij} \sim Poisson(\cdot|\sum_{k}\mathbf V_{ik}\mathbf V_{jk})$;
	\item For each vertex $i$, draw $\mathbf F_{ji}\!\sim\! \lbrack\sum_{k}\mathbf V_{ik}\mathbf U_{jk}\rbrack^{\mathbf F_{ji}}$; draw $\mathbf V_{ik}, \sum_{l}\mathbf X_{il}\mathbf H_{lk}\sim\mathcal{N}(\mathbf V_{ik}-\sum_{l}\mathbf X_{il}\mathbf H_{lk}|0,\lambda_i)$;
\end{itemize}
where $\lambda_i$, and $C$ are the precision of Gaussian distribution, and product of Boltzmann constant and temperature, respectively.
Given such a generative process, it is found that RTCMPN is fundamentally different from previous models for network clustering.
Besides considering modeling edge structure and cluster themes utilizing vertex-cluster membership, RTCMPN further learns the latent neighboring preference ($\mathbf X$), according to the adaptive dichotomous inclinations ($\mathbf S$) and corresponding observed similarities ($\mathbf Z$ and $\mathbf G$).
Such neighbor preference is used to regularize the vertex-cluster preference via the estimated cluster-vertex proportions, so that pairwise vertices possessing similar topological/feature inclinations and local neighbors are more possible to be assigned with similar cluster membership.
RTCMPN is also different from previous approaches to learning adaptive neighbors~\cite{nie2014clustering,wang2019gmc,wang2019parameter}, as they are neither capable of modeling vertex-wise topological/feature inclinations, nor designed for network clustering.
By making use of RTCMPN, one can acquire more descriptive information in the network, including latent cluster structure, cluster themes, vertex-wise topological/feature inclination indicating how the vertex is related to others, and the neighboring structure within each cluster.

\subsection{The joint likelihood}
Based on the generative process introduced in Section~\ref{structure}, the joint likelihood of the proposed model is:
\begin{equation}
\label{jointlikeli}
\begin{aligned}
&p(\mathbf Y,\mathbf F,\mathbf Z,\mathbf G,\mathbf V,\mathbf U,\mathbf X,\mathbf H,\mathbf S|\lambda)=\\
&\prod_{ i \ne j} p(\mathbf Y_{ij}|\sum_{k}\mathbf V_{ik}\mathbf V_{jk})\cdot\prod_{i, k} \!p(\mathbf V_{ik},\sum_{l}\mathbf X_{il}\mathbf H_{lk}|\lambda_i)\\
&\cdot\!\prod_{i, j} \!p(\mathbf F_{ji}|\sum_{k}\mathbf V_{ik}\mathbf U_{jk})\!\cdot\!\prod_{ i \ne j} \!p(\mathbf X_{ij},\mathbf S_{i},\mathbf S_{j}, \mathbf Z_{ij}, \mathbf G_{ij}).\\
\end{aligned}
\end{equation}
Taking the logarithm of the joint likelihood and substituting the aggregation operators with corresponding matrix operators, we have:
\begin{equation}
\label{loglikeli}
\begin{aligned}
&L(\mathbf Y,\mathbf F,\mathbf Z, \mathbf G, \mathbf V,\mathbf U,\mathbf X,\mathbf H,\mathbf S|\lambda) =K\sum_{i}\!\log\sqrt{\lambda_i}\\
&-\!\sum_{i,k}\frac{\lambda_i}{2}\lbrack\mathbf V_{ik}-(\mathbf X\mathbf H)_{ik}\rbrack^2\!+\!\sum_{i,j}\mathbf F_{ji}\!\log(\mathbf U\mathbf V^T)_{ji}\\
&\sum_{i \ne j}\!\lbrack\mathbf Y_{ij}\!\log(\mathbf V\mathbf V^T)_{ij}\!-\!(\mathbf V\mathbf V^T)_{ij}\rbrack-\log A\\
&+ \sum_{i \ne j}\lbrack\mathbf X_{ij}(\mathbf S_{i1}\mathbf S_{j1}\mathbf Z_{ij}+\mathbf S_{i2}\mathbf S_{j2}\mathbf G_{ij})\rbrack+const,
\end{aligned}
\end{equation}
where $A$ is the normalization term for Boltzmann distribution, which is $\sum_{i\ne j}\exp\{-\epsilon_{ij}\}$, and $const$ contains the terms that are irrelevant to all the latent variables.
Based on Eq.~(\ref{loglikeli}), we can observe that the joint likelihood of the proposed model may increase when the edges are generated by appropriate cluster preferences of the corresponding bridging vertices, the features are appropriate cluster themes, and preferred neighbors are in the same cluster. 
Therefore, the vertex-cluster preferences which we expect RTCMPN to learn can be achieved when Eq.~(\ref{loglikeli}) is maximized.

\subsection{Learning RTCMPN}
As Eq.~(\ref{loglikeli}) shows, it is intractable to directly optimize the model through point estimation since the summation operators are embedded into the logarithm operator.
As a result, the latent variables, including $\mathbf V$ and $\mathbf U$, can only be optimized via approximation methods. 
In this paper, we derive an alternative manner for updating the latent variables of the model, based on Expectation-Maximization (EM) framework~\cite{dempster1977}.
In the E-step of each iteration, RTCMPN constructs an auxiliary function which manifests the lower bound of the log-likelihood function regarding the latent variables.
Then, this lower bound is maximized by RTCMPN in the M-step.
By iteratively updating the latent variables using EM algorithm, the proposed model can converge in a finite number of iterations.
\subsubsection{E-step}
To derive the lower bound of Eq.~(\ref{loglikeli}) in E-step, we use the following well known property of the concavity of logarithmic functions:
\begin{equation}
\begin{aligned}
&\log(\sum_{k}\!x_k)\!\ge\!\sum_{k}a_k\log(\frac{x_k}{a_k}),\text{if }\sum_{k}a_k \!= \!1, x_k>0.
\end{aligned}
\end{equation}
Based on it, we may derive the following auxiliary function as the lower bound of the joint likelihood of RTCMPN (Eq.~(\ref{loglikeli})):
\begin{equation}
\label{lowerbound}
\begin{aligned}
&L(\mathbf Y,\mathbf F,\mathbf Z, \mathbf G, \mathbf V,\mathbf U,\mathbf X,\mathbf H,\mathbf S|\lambda)\ge Q(\theta, \phi)=\\
&\sum_{i\ne j}\lbrack\mathbf Y_{ij}\sum_{k}\theta_{ij,k}\log(\frac{\mathbf V_{ik}\mathbf V_{jk}}{\theta_{ij,k}})-\!(\mathbf V\mathbf V^T)_{ij}\rbrack\\
&+\sum_{i,j}\lbrack\mathbf F_{ji}\sum_{k}\phi_{ji,k}\log(\frac{\mathbf V_{ik}\mathbf U_{jk}}{\phi_{ji,k}})\rbrack+\sum_{i}K\log\sqrt{\lambda_i}\\
&-\sum_{i,k}\!\frac{\lambda_i}{2}\lbrack\mathbf V_{ik}^2-2(\mathbf {XH})_{ik}\mathbf V_{ik}+(\mathbf {XH})^2_{ik}\rbrack-\log A\\
&+\sum_{i\ne j}\mathbf {X}_{ij}(\mathbf {S}_{i1}\mathbf {S}_{j1}\mathbf {Z}_{ij}+\mathbf {S}_{i2}\mathbf {S}_{j2}\mathbf {G}_{ij})+const,\\
&\theta_{ij,k} = \frac{\mathbf V_{ik}\mathbf V_{jk}}{\sum_{k}\mathbf V_{ik}\mathbf V_{jk}},\phi_{ji,k}=\frac{\mathbf V_{ik}\mathbf U_{jk}}{\sum_{k}\mathbf V_{ik}\mathbf U_{jk}}.\\
\end{aligned}
\end{equation}
In E-step, we set the plug-in variables, including $\theta_{ij,k}$ and $\phi_{ji,k}$ as Eq.~(\ref{lowerbound}) shows, so that these plug-in variables directly take effect on the corresponding latent variables of the model.
Then, we are able to maximize Eq.~(\ref{loglikeli}) via pushing-up the lower bound $Q(\theta, \phi)$.

\subsubsection{M-step}

In M-step, we maximize the lower bound $Q(\theta, \phi)$, which is shown in Eq.~(\ref{lowerbound}), by simply performing point estimation. To optimize $Q$ relevant to latent variable $\mathbf V_{ik}$, we construct the following Lagrangian function:
\begin{equation}
\label{lagrange}
\mathcal L(\mathbf V_{ik},\nu) = Q(\theta, \phi) - \nu\lbrack\sum_{k}\mathbf V_{ik}-1\rbrack,
\end{equation}
where $\nu$ denotes the Lagrange multiplier in terms of the unity constraint of $\mathbf V_{i\cdot}$. Letting the partial derivative w.r.t. $\mathbf V_{ik}$ of Eq.~(\ref{lagrange}) be equal to zero and substituting $\nu$, we may derive the updating rule for $\mathbf V_{ik}$:
\begin{equation}
\label{max-v}
\begin{aligned}
&\mathbf V_{ik} =\frac{\Delta_{ik}\sum_k\frac{\mathbf V_{ik}}{\Lambda_{ik}}+\mathbf V_{ik}}{\Lambda_{ik}\sum_k\frac{\mathbf V_{ik}}{\Lambda_{ik}}+\sum_k\frac{\mathbf V_{ik}\Delta_{ik}}{\Lambda_{ik}}},\\
&\Delta_{ik} \!=\! 2\sum_{j}\mathbf Y_{ij}\theta_{ij,k}+\sum_{j}\mathbf F_{ji}\phi_{ji,k}+\lambda_i(\mathbf{XH})_{ik}\mathbf V_{ik},\\
&\Lambda_{ik} = 2\sum_{j}\mathbf V_{jk}+\lambda_i\mathbf V_{ik}.\\
\end{aligned}
\end{equation}
Similarly, we may obtain the updating rules for $\mathbf U_{jk}$, $\mathbf X_{ij}$, $\mathbf H_{jk}$, and $\mathbf S_{id}$ for $d\in\{1,2\}$.
The rule for updating $\mathbf U_{jk}$ is:
\begin{equation}
\label{max-u}
\begin{aligned}
\mathbf U_{jk} = \frac{\sum_{i}\mathbf F_{ji}\phi_{ji,k}}{\sum_{i,j}\mathbf F_{ji}\phi_{ji,k}}.
\end{aligned}
\end{equation}
The updating rule for $\mathbf X_{ij}$ is:
\begin{equation}
\label{max-x}
\begin{aligned}
&\mathbf X_{ij} =\frac{\Delta_{ij}\sum_j\frac{\mathbf X_{ij}}{\Lambda_{ij}}+\mathbf X_{ij}}{\Lambda_{ik}\sum_j\frac{\mathbf X_{ij}}{\Lambda_{ij}}+\sum_j\frac{\mathbf X_{ij}\Delta_{ij}}{\Lambda_{ij}}},\\
&\Delta_{ij}\! =\! \lbrack\!\lambda_i(\!\mathbf {V}\!\mathbf {H}^T\!)_{ij}\!+\!\eta_{ij}\!\rbrack\mathbf X_{ij}\!,\!\eta_{ij}\! =\! \mathbf S_{i1}\!\mathbf S_{j1}\!\mathbf Z_{ij}\!+\!\mathbf S_{i2}\!\mathbf S_{j2}\!\mathbf G_{ij},\\
&\Lambda_{ij} \!=\! \lambda_i(\mathbf {X}\mathbf {HH}^T)_{ij}\!+\!\eta_{ij}\frac{\exp\{-\epsilon_{ij}\}}{A}.\\
\end{aligned}
\end{equation}
The updating rule for $\mathbf H_{jk}$ is:
\begin{equation}
\label{max-h}
\begin{aligned}
&\mathbf H_{jk} =\frac{\mathbf H_{jk}(\mathbf {XV})_{jk}\sum_j\frac{\mathbf H_{jk}}{(\mathbf {X}^T\mathbf {XH})_{jk}}+\mathbf H_{jk}}{(\mathbf {X}^T\mathbf {XH})_{jk}\sum_j\frac{\mathbf H_{jk}}{(\mathbf {X}^T\mathbf {XH})_{jk}}+\sum_j\frac{\mathbf H_{jk}(\mathbf {XV})_{jk}}{(\mathbf {X}^T\mathbf {XH})_{jk}}}.\\
\end{aligned}
\end{equation}
The updating rule for $\mathbf S_{id}$ is:
\begin{equation}
\label{max-s}
\begin{aligned}
&\mathbf S_{i1} \!=\!\frac{\mathbf S_{i1}(\frac{\mathbf S_{i1}}{(\Psi\mathbf S)_{i1}}+\frac{\mathbf S_{i2}}{(\Phi\mathbf S)_{i2}})(\Delta\mathbf S)_{i1}+\mathbf S_{i1}}{\!(\!\Psi\mathbf S\!)_{i1}\!(\frac{\mathbf S_{i1}}{\!(\Psi\mathbf S)_{i1}}\!+\!\frac{\mathbf S_{i2}}{\!(\Phi\mathbf S)_{i2}})\!+\!(\frac{\mathbf S_{i1}\!(\Delta\mathbf S)_{i1}}{\!(\Psi\mathbf S)_{i1}}\!+\!\frac{\mathbf S_{i2}\!(\!\Lambda\mathbf S\!)_{i2}}{\!(\Phi\mathbf S)_{i2}})},\\
&\mathbf S_{i2}\!=\!\frac{\mathbf S_{i2}(\frac{\mathbf S_{i1}}{(\Psi\mathbf S)_{i1}}+\frac{\mathbf S_{i2}}{(\Phi\mathbf S)_{i2}})(\Lambda\mathbf S)_{i2}+\mathbf S_{i2}}{\!(\Phi\mathbf S)_{i2}\!(\frac{\mathbf S_{i1}}{\!(\Psi\mathbf S)_{i1}}\!+\!\frac{\mathbf S_{i2}}{\!(\!\Phi\mathbf S)_{i2}}\!)\!+\!(\frac{\mathbf S_{i1}\!(\Delta\mathbf S)_{i1}}{\!(\Psi\mathbf S)_{i1}}\!+\!\frac{\mathbf S_{i2}\!(\Lambda\mathbf S)_{i2}}{\!(\Phi\mathbf S)_{i2}})}\!,\\
&\!\Delta_{ij} = \mathbf{X}_{ij}\mathbf{Z}_{ij},\Lambda_{ij} = \mathbf{X}_{ij}\mathbf{G}_{ij},\\
&\Psi_{ij}\! =\! \mathbf{X}_{ij}\mathbf{Z}_{ij}\frac{\exp\{-\epsilon_{ij}\}}{A}, \Phi_{ij}\! =\! \mathbf{X}_{ij}\mathbf{G}_{ij}\frac{\exp\{-\epsilon_{ij}\}}{A}.
\end{aligned}
\end{equation}
Finally, $\lambda_i$ can be updated through performing MLE:
\begin{equation}
\label{max-lambda}
\begin{aligned}
\lambda_i = \frac{K}{\sum_{k}\lbrack\mathbf V_{ik}-(\mathbf {XH})_{ik}\rbrack^2}.
\end{aligned}
\end{equation}
By iteratively performing E-step and M-step, RTCMPN will converge to local optima in a finite number of iterations. The process of variable learning of RTCMPN has been summarized in Algorithm~1.

\begin{algorithm}
	\caption{Relational thematic clustering with mutually preferred neighbors (RTCMPN)}
	\begin{algorithmic}[1]
		\REQUIRE{Network Data: $\mathbf Y$, $\mathbf F$, $\mathbf Z$, $\mathbf G$}
		\ENSURE{Cluster preference for each vertex $\mathbf \lbrace \mathbf V_{i\cdot}\rbrace_{i=1}^N$; Cluster themes $\mathbf \lbrace \mathbf U_k\rbrace_{k=1}^K$; Vertex-cluster proportion $\mathbf \lbrace \mathbf H_{k}\rbrace_{k=1}^K$; Latent neighborhood preference $\mathbf \lbrace \mathbf X_{ij}\rbrace_{i,j=1}^N$; Vertex-wise topology/feature inclination $\mathbf \lbrace \mathbf S_{i\cdot}\rbrace_{i=1}^N$}\\
		
		\STATE Initialize $\mathbf U$, $\mathbf V$, $\mathbf X$,  $\mathbf S$, $\mathbf H$, $\lambda_i$;\\
		
		\STATE $t \leftarrow 0$;\\
		\WHILE{$t < T_{max}$}
		\STATE $t \leftarrow t+1$;\\
		\STATE E-Step: Set lower bound of Eq.~(\ref{loglikeli}) by Eq.~(\ref{lowerbound});
		\STATE M-Step: Maximize the lower bound via:
		\STATE Updating $\mathbf V_{ik}$, $\mathbf U_{jk}$, $\mathbf X_{ij}$, $\mathbf H_{jk}$, $\mathbf S_{id}$, and $\lambda_{i}$ by Eqs. (\ref{max-v})-(\ref{max-lambda});\\
		
		\STATE Compute Log likelihood $L^{(t)}$ by Eq.~(\ref{loglikeli});\\
		\IF{$L^{(t)}-L^{(t-1)}\le \epsilon$}\STATE break;\\ \ENDIF
		\ENDWHILE
		\STATE Identify cluster label for each vertex using $\mathbf V$.\\
		
		\label{RTCMPN-learning}
	\end{algorithmic}
\end{algorithm}

\subsection{Complexity analysis}
Based on the E-step and M-step shown in Eq.~(\ref{lowerbound}), and Eqs.~(\ref{max-v})-(\ref{max-lambda}), the complexity of the proposed model can be approximately analyzed as follows.
In E-step, setting lower-bounder for $\mathbf V_{ik}$, and $\mathbf U_{jk}$ follows the order of $O(K)$.
In M-step, updating $\mathbf V_{ik}$ and $\mathbf U_{jk}$ follows the order of $O(3N+M)$, and $O(2N)$, respectively.
For $\mathbf X_{ij}$, RTCMPN considers only those connected vertices to improve the computational efficiency.
Thus, updating $\mathbf X_{ij}$, $\mathbf H_{jk}$, $\mathbf S_{id}$, and $\mathbf \lambda_{i}$ follows the order of $O((e+2)K)$, $O((N+2)e)$, $O(4e)$, and $O((e+1)K)$, where $e$ represents the average vertex degree.

\begin{table}
	\centering
	\resizebox{\linewidth}{!}{
	\begin{tabular}{llrrrr}  
		\toprule
		Dataset     & Type & $N$      & $|E|$  & $M$     & $K$    \\
		\midrule
		Ego-facebook         & Soc  & 4039   & 88234  & 1283  & 191  \\
		Google+       & Soc  & 107614 & 3755989& 13966 & 463  \\
		Washington        & Doc  & 230    & 366    & 1579  & 5    \\
		UAI         & Doc  & 3363   & 33300  & 4971  & 19   \\
		Wiki        & Doc  & 2405   & 17981  & 4973  & 17   \\
		Biogrid     & Bio  & 5640   & 59748  & 4286  & 200  \\
		\bottomrule
	\end{tabular}
	}
	\caption{Statistics of datasets used in the experiments. Soc, Doc, or Bio represents whether the dataset is a social, document, or biological network.}
	\label{dataset}
\end{table}

\begin{table*}
	\centering
	\resizebox{\linewidth}{40mm}{
		\begin{tabular}{lrrrrrrrrrrrrrr}
			\toprule
			Dataset   & \multicolumn{2}{c}{Ego-facebook}   & \multicolumn{2}{c}{Google+}   & \multicolumn{2}{c}{Washington}    & \multicolumn{2}{c}{UAI}   & \multicolumn{2}{c}{Wiki}    & \multicolumn{2}{c}{Biogrid} \\  
			\midrule                        
			Metrics & $NMI$ & $Acc$ & $NMI$ & $Acc$ & $NMI$ & $Acc$ & $NMI$ & $Acc$ & $NMI$ & $Acc$ & $NMI$ & $Acc$ \\
			\midrule                        
			\multirow{2}{*}{Ncut} & 53.646 &  44.689 &  5.122 & 14.921 &  16.151 &  63.043 &  16.855 &  29.527 &  8.638 & 17.588 &  83.458 &  3.245 \\
			& $\pm$0.162 &  $\pm$0.520 &  $\pm$0.587 &  $\pm$0.188 &  $\pm$0.981 &  $\pm$0.652 &  $\pm$0.272 &  $\pm$0.164 &  $\pm$0.412 &  $\pm$0.208 &  $\pm$0.262 &  $\pm$0.098 \\
			\hline                        
			\multirow{2}{*}{SBM} &  63.527 &  52.959 &  9.871 & 27.669 &  12.838 &  46.522 &  28.693 &  43.622 &  22.670 &  36.715 &  88.868 &  8.528 \\
			& $\pm$0.403 &  $\pm$0.186 &  $\pm$0.437 &  $\pm$0.175 &  $\pm$0.007 &  $\pm$0.048 &  $\pm$0.276 &  $\pm$0.104 &  $\pm$0.008 &  $\pm$0.748 &  $\pm$0.734 &  $\pm$0.363 \\
			\hline                        
			\multirow{2}{*}{M-NMF} &  26.428 &  33.147 &  6.315 & 17.269 &  18.019 &  46.522 &  17.160 &  25.968 &  17.168 &  36.923 &  81.938 &  4.078 \\
			& $\pm$1.646 &  $\pm$0.944 &  $\pm$0.753 &  $\pm$1.198 &  $\pm$1.366 &  $\pm$0.913 &  $\pm$1.422 &  $\pm$1.624 &  $\pm$1.064 &  $\pm$1.759 &  $\pm$0.988 &  $\pm$0.635 \\
			\hline                        
			\multirow{2}{*}{DANMF} &  56.447 &  43.352 &  10.423 &  13.723 &  13.225 &  50.233 &  31.812 &  43.816 &  33.889 &  50.148 &  70.885 &  8.865 \\
			& $\pm$0.601 &  $\pm$1.493 &  $\pm$1.685 &  $\pm$0.631 &  $\pm$0.541 &  $\pm$1.395 &  $\pm$1.497 &  $\pm$0.834 &  $\pm$0.803 &  $\pm$1.201 &  $\pm$0.468 &  $\pm$0.227 \\
			\hline                        
			\multirow{2}{*}{$k$-means} &  40.461 &  29.116 &  6.421 & 10.385 &  26.738 &  68.043 &  33.947 &  42.284 &  33.972 &  41.435 &  89.163 &  8.475 \\
			& $\pm$0.818 &  $\pm$0.248 &  $\pm$0.572 &  $\pm$0.036 &  $\pm$3.936 &  $\pm$2.826 &  $\pm$3.402 &  $\pm$3.613 &  $\pm$1.947 &  $\pm$0.478 &  $\pm$0.454 &  $\pm$0.284 \\
			\hline                        
			\multirow{2}{*}{MVSC} & 12.710 &  14.236 &  7.580 & 14.192 &  12.737 &  47.826 &  13.624 &  27.059 &  16.736 &  19.958 &  77.401 &  4.609 \\
			& $\pm$0.318 &  $\pm$0.272 &  $\pm$0.983 &  $\pm$0.961 &  $\pm$0.279 &  $\pm$0.001 &  $\pm$0.190 &  $\pm$0.327 &  $\pm$0.306 &  $\pm$0.187 &  $\pm$0.249 &  $\pm$0.098 \\
			\hline                        
			\multirow{2}{*}{CESNA} &  57.513 &  46.124 &  10.164 &  26.783 &  22.177 &  64.929 &  17.536 &  23.839 &  9.374 & 24.738 &  80.088 &  2.570 \\
			& $\pm$1.119 &  $\pm$1.118 &  $\pm$0.601 &  $\pm$0.441 &  $\pm$0.536 &  $\pm$0.664 &  $\pm$0.566 &  $\pm$0.570 &  $\pm$0.185 &  $\pm$0.144 &  $\pm$0.353 &  $\pm$0.112 \\
			\hline                        
			\multirow{2}{*}{SCI} &  44.025 &  35.113 &  10.241 &  15.812 &  6.912 & 53.043 &  29.673 &  44.643 &  22.012 &  38.843 &  \textbf{92.028} &  \textbf{12.720} \\
			& $\pm$0.509 &  $\pm$0.124 &  $\pm$0.775 &  $\pm$0.065 &  $\pm$0.680 &  $\pm$0.435 &  $\pm$1.397 &  $\pm$1.091 &  $\pm$0.536 &  $\pm$0.108 &  $\pm$0.195 &  $\pm$0.063 \\
			\hline                        
			\multirow{2}{*}{ASCD} & 45.857 &  35.746 &  6.704 & 17.505 &  19.967 &  59.908 &  24.688 &  30.370 &  26.965 &  32.952 &  86.296 &  6.986 \\
			& $\pm$0.249 &  $\pm$0.087 &  $\pm$1.232 &  $\pm$0.578 &  $\pm$0.637 &  $\pm$5.760 &  $\pm$1.256 &  $\pm$0.504 &  $\pm$1.256 &  $\pm$0.083 &  $\pm$0.314 &  $\pm$0.124 \\
			\hline
			\multirow{2}{*}{GMC} & 55.770 &  44.194 &  9.435 & 21.143 &  23.962 &  66.522 &  34.009 &  30.865 &  18.788 &  31.102 &  70.564 &  7.092 \\
			& $\pm$0 &  $\pm$0 &  $\pm$0 &  $\pm$0 &  $\pm$0 &  $\pm$0 &  $\pm$0 &  $\pm$0 &  $\pm$0 &  $\pm$0 &  $\pm$0 &  $\pm$0 \\
			\hline                        
			\multirow{2}{*}{RTCMPN} & \textbf{68.836} &  \textbf{59.599} &  \textbf{16.112} &  \textbf{58.628} &  \textbf{30.138} &  \textbf{69.565} &  \textbf{39.686} &  \textbf{52.156} &  \textbf{49.447} &  \textbf{62.453} &  91.116 &  12.163 \\
			& $\pm$0.115 &  $\pm$0.301 &  $\pm$0.256 &  $\pm$0.064 &  $\pm$2.387 &  $\pm$2.574 &  $\pm$1.426 &  $\pm$0.921 &  $\pm$1.045 &  $\pm$1.672 &  $\pm$0.164 &  $\pm$0.333 \\
			\bottomrule
	\end{tabular}}
	\caption{Clustering Performance Evaluated by $NMI$ and $Acc$ (mean $\pm$ std. deviation). The best performance on each dataset is highlighted in bold.}
	\label{performance}
\end{table*}

\section{Experimental analysis}\label{exp}
In this section, we conduct a series of experiments on real-world datasets, including social network, document network, and biological network to validate the effectiveness of RTCMPN against state-of-the-art methods.

\subsection {Experimental setup}
\subsubsection{Baselines for comparison}
Ten approaches are selected as baselines, which can be categorized into three classes. Ncut \cite{shi2000}, SBM \cite{peng2015scalable}, M-NMF \cite{wang2017community}, and DANMF \cite{ye2018deep} are four prevalent methods utilizing network topology to uncover clusters. $k$-means \cite{mackay2003information} is an effective vertex-feature-based approach to network clustering. MVSC \cite{kumar2011co}, CESNA \cite{yang2013}, SCI \cite{wang2016semantic}, ASCD \cite{qin2018adaptive}, and GMC~\cite{wang2019gmc} are state-of-the-art approaches to network clustering, which utilize both network structure and vertex features to unfold clusters in the network.

In our experiments, we used the source codes of all the baselines provided by the authors for implementation and configured the baselines by using either default settings or recommended ones.
Specifically, CESNA does not need any predefined parameter.
For the rest of the baselines, including SBM, M-NMF, DANMF, $k$-means, MVSC, CESNA, SCI, ASCD, and GMC, we used the default settings recommended by the authors.
For the number of clusters, i.e., $K$, which has to be determined by all the baselines except CESNA, and the proposed model, we set it to be equal to the number of ground-truth clusters of the testing dataset.
All of the experiments were performed on a workstation with 6-core 3.4GHz CPU and 32GB RAM and all approaches were executed 10 times to obtain a statistically steady performance.

\subsubsection{Dataset description}
We used six real-world networks with verified ground-truth clusters as testing datasets, including two social graphs, three document networks, and one biological network.
These real-world networks have different sizes and different numbers of vertex features. 
$Ego-facebook$ (Ego)~\cite{leskovec2012learning} and $Google+$ (Gplus)~\cite{mcauley2014discovering} are two social networks whose vertices and edges respectively represent the social networking users and their social relationships. 
$Washington$ (Wash)~\cite{lu2003}, $UAI$, and $Wiki$~\cite{lu2003} are three widely used document networks, whose vertices and edges represent the documents, and the citations/hyperlinks between pairwise documents, respectively.
$Biogrid$~\cite{stark2006biogrid} is a biological network used to describe the interactions between proteins in $Saccharomyces$ $cerevisiae$.
The statistics of these testing datasets are summarized in Table~\ref{dataset}, where $N$ is number of vertices, $|E|$ is the number of edges, $M$ is the number of vertex features, and $K$ is the number of ground-truth clusters, respectively.

\subsubsection {Evaluation metrics}
Two prevalent metrics, that are Normalized Mutual Information ($NMI$)~\cite{he2018} and the Accuracy ($Acc$)~\cite{he2019end}, are used in our experiments to evaluate the performance of different approaches.
According to their definitions, larger values of $NMI$ and $Acc$ indicate a better matching between the detected clusters and the ground-truth.

\subsection {Clustering performance comparison}
Social community detection, document segmentation, and biological module identification are typical applications of network clustering.
In our experiment, we used the aforementioned networks to test the effectiveness of different approaches.
The experimental results (in terms of $NMI$ and $Acc$) of all algorithms are summarized in Table \ref{performance}.

When the detected clusters are evaluated by $NMI$, RTCMPN outperforms all the other baselines in five testing datasets, and ranks the second-best on $Biogrid$ dataset.
In five datasets out of six, RTCMPN is better than the second best approach by at least 5\%.
Specifically, in $Ego-facebook$ dataset, RTCMPN performs better than SBM by 8.36\%.
In $Google+$, RTCMPN outperforms DANMF by 54.58\%. In $Washington$, and $Wiki$, RTCMPN is better than $k$-means by 12.72\%, and 45.55\%, respectively. In $UAI$, the proposed model outperforms GMC by 16.69\%.

When $Acc$ is considered, RTCMPN still performs robustly when compared with other baselines. RTCMPN can obtain the best performance in terms of $Acc$ in all the testing datasets, except $Biogrid$, where it ranks the second best.
RTCMPN may outperform the second best approaches by at least 10\% in four datasets.
In $Ego-facebook$ and $Google+$, RTCMPN outperforms SBM by 12.54\% and 111.89\%, respectively. In $UAI$, RTCMPN is better than SCI by 16.83\%.
In $Wiki$, RTCMPN outperforms DANMF by 24.54\%.

From the experimental results in terms of $NMI$ and $Acc$, we can observe that RTCMPN is effective in network clustering.
It is the novel model structure that makes the proposed approach outperform other baselines.
Considering modeling the latent neighboring preference which is also aware of topology/feature inclinations of each vertex, RTCMPN is able to assign similar cluster membership to those vertices having analogous structure of latent local neighbors and preference in terms of topology and feature.
More meaningful clusters can thereby be uncovered by the proposed model.

\begin{figure}
	\centering
	\includegraphics[width=\linewidth]{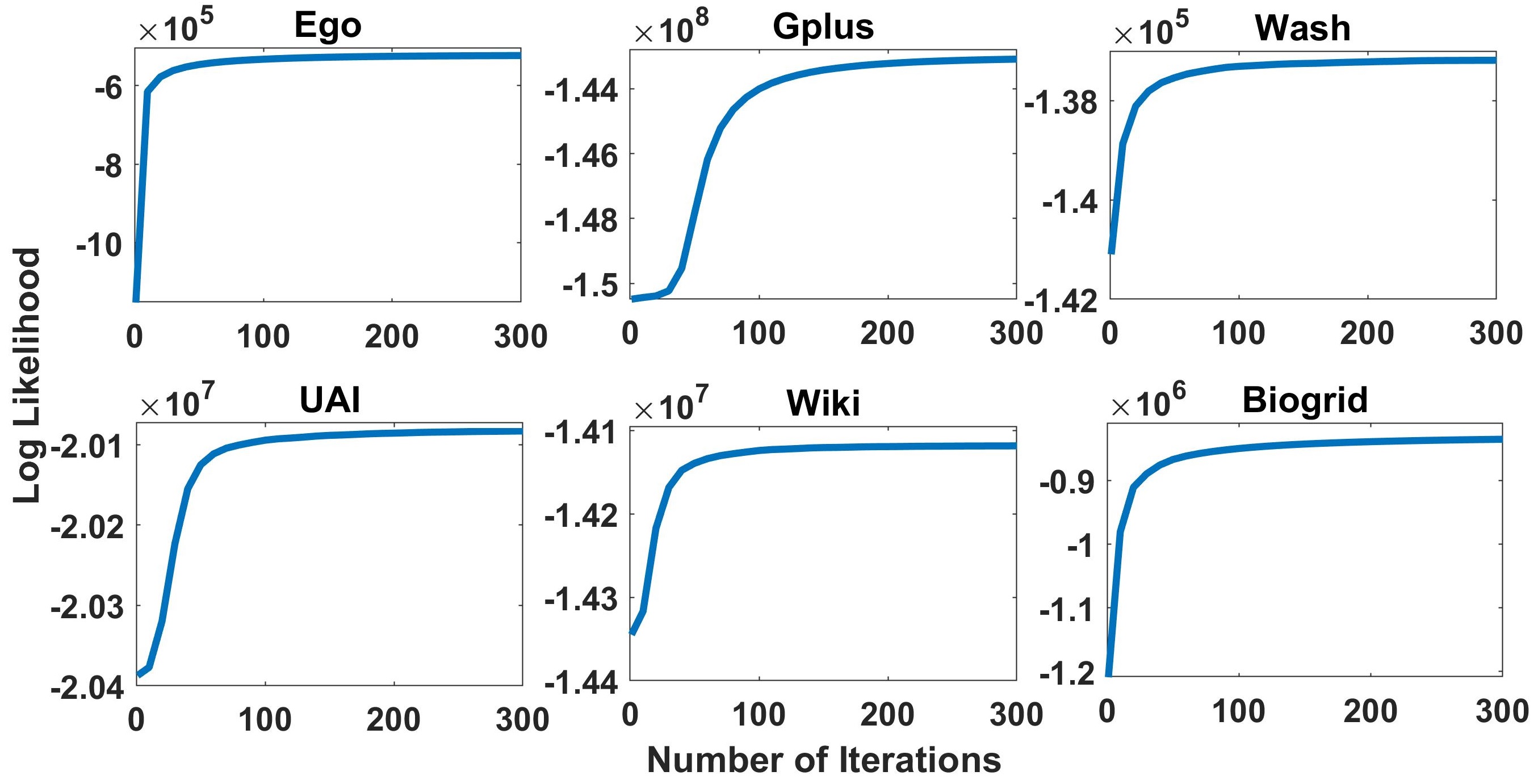}
	\caption{Model convergence in testing datasets}
	\label{conv}
\end{figure}

\subsection{Model convergence test}
In addition to derive the EM algorithm for updating the latent variables of RTCMPN, we also investigated the convergence speed of the proposed method on real network datasets.
Specifically, we recorded the value of log-likelihood function for the first 300 iterations on all the six datasets. 
As depicted in Fig.~\ref{conv}, the value of log-likelihood converges to a stable value in finite iterations, which showcases the capability of the derived EM algorithm to guarantee the model convergence and attain the optimal clustering results efficiently.


\begin{figure}
	\centering
	\tikzstyle{mycir} = [circle, draw, text centered]
	\tikzstyle{myrec} = [rectangle, draw]
	\tikzset{global scale/.style={transform shape, scale=1}}
	\tikzstyle{every node}=[font=\scriptsize]
	\tikzset{>=latex}
	\centering
	\resizebox{1\linewidth}{!}{
	\begin{tikzpicture}[global scale,->,>=stealth']
	
	\node [mycir] (n1) at (0,0) {\textbf{58}};
	\node [myrec] (n2) at (3,0.3) {${\textbf X}_{(58,2814)}$: 0.139};
	\node [myrec] (n3) at (3,-0.3) {${\textbf X}_{(2814,58)}$: 0.7801};
	\node [mycir] (n4) at (6,0) {\textbf{2814}};
    \node [myrec,align=center] (n5) at (0,-0.9) {${\textbf S}_{(58,1)}$:0.5303\\ ${\textbf S}_{(58,2)}$:0.4697};
    \node [myrec,align=center] (n6) at (5.9,-0.9) {${\textbf S}_{(2814,1)}$:0.4622\\ ${\textbf S}_{(2814,2)}$:0.5378};
    \node [myrec,align=left] (n7) at (0, -2.7) {\textbf{58-Features} \\ Anonymous 1 \\ Anonymous 3\\ \textit{Anonymous 10}\\ \textit{Anonymous 25}\\ Anonymous 51\\ Anonymous 68\\ \textit {\textbf {Anonymous 115}}\\ \textbf{Anonymous 116}};
    \node [myrec,align=left] (n8) at (3, -1.7) {\textbf{Cluster Theme} \\ \textbf{Anonymous 106} \\ \textbf{Anonymous 115}\\ \textbf{Anonymous 116}\\ Anonymous 185\\ Anonymous 620};
    \node [myrec,align=left] (n9) at (5.9, -2.58) {\textbf{2814-Features} \\ \textit{Anonymous 10} \\ \textit{Anonymous 25}\\Anonymous 39\\ \textbf{Anonymous 106}\\ \textit{\textbf{Anonymous 115}}\\ Anonymous 359\\ Anonymous 1090};

    \node [align = left] (n10) at (3.1, -3.3) { \textbf{Boldface}: Feature shared \\ by cluster and vertex \\ \textit{Italic}: Feature shared \\ by vertices};

    \draw[-] (-0.9, -1.7)--(0.9, -1.7);
    \draw[-] (5.2, -1.7)--(6.8, -1.7);
    \draw[-] (2.1, -1.1)--(3.9, -1.1);
    \draw[dashed] (n1)--(n2);
    \draw[dashed] (n3)--(n1);
    \draw[dashed] (n2)--(n4);
    \draw[dashed] (n4)--(n3);
	\end{tikzpicture}}
	\caption{Mutually preferred neighbors in $Ego-facebook$ dataset.}\label{neighbor}
\end{figure}
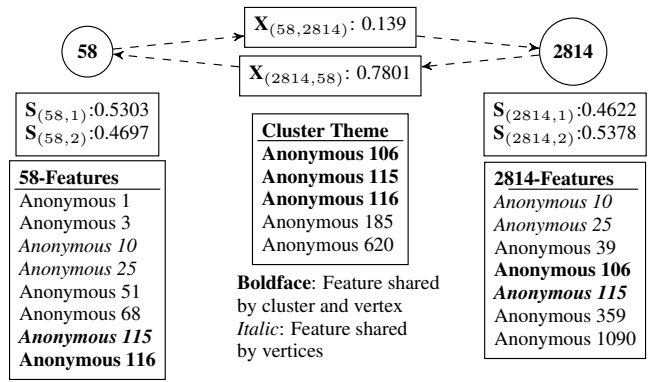

\subsection{Scalability comparison}
To show the scalability of RTCMPN, we compared the computational time of RTCMPN with CESNA, which is a well known efficient model-based approach to network clustering.
The optimization time used by RTCMPN in datasets $Ego-facebook$, $Google+$, $Washington$, $UAI$, $Wiki$, and $Biogrid$ are 246.43, 12756.90, 1.95, 212.87, 131.89, and 585.99 seconds, respectively.
While, the optimization time used by CESNA in the corresponding datasets are 372, 19320, 12.250, 850, 342, and 4080 seconds, respectively.
It can be seen that the efficiency of RTCMPN outperforms CESNA.

\subsection{Case study on mutually preferred neighbors}
To verify whether the proposed model can uncover clusters by considering the mentioned latent preference and dichotomous inclinations, we conducted a detailed analysis on the clusters discovered by RTCMPN and provide a concrete example.
Figure~\ref{neighbor} illustrates a pair of cluster members correctly detected by RTCMPN in $Ego-facebook$ dataset.
Why they are correctly detected in the same cluster can be explained using the depicted latent information learned by RTCMPN.
It is observed that the mutual neighboring preferences between vertex 58 and 2814 are both larger than 0.1, which is a relatively high value in view of the size of the dataset ($N$=4039).
As the topology/feature inclinations of these two vertices are very similar, RTCMPN deduces they are very possible to be associated (mutually preferred) and consequently learns high neighboring preferences.
The clustering performance of RTCMPN is thereby improved by assigning such mutually preferred vertices with similar cluster membership.

\section*{Conclusion}
In this paper, we propose a novel probabilistic model for network clustering, dubbed Relational Thematic Clustering with Mutually Preferred Neighbors (RTCMPN).
Different from previous approaches which mainly utilize network topology and vertex features to unfold clusters in the network, RTCMPN further learns the latent preferences indicating which vertices and their neighbors are more possible to be in the same cluster, according to the vertex-wise dichotomous inclinations w.r.t. topology and features.
Such latent neighboring preferences are then used to guide the proposed model to assign more analogous cluster membership to those vertices having similar proximal preferences and topology/feature inclinations.
More interpretable clusters can thereby be learned by the proposed model.
Having been compared with a number of strong baselines on various types of networks, RTCMPN is found to be more effective in unfolding network clusters.
In future, we will further improve the interpretability of RTCMPN via developing the fully Bayesian version of the model and enhance its efficiency by deriving stochastic learning algorithms for model inference. 

{
	\balance
	\bibliographystyle{named}
	\bibliography{RTCMPN}
}

\end{document}